\documentstyle[aps,twocolumn,epsf]{revtex}
\begin{document} 

\draft

\preprint{ZHONG Fan et al}

\twocolumn[\hsize\textwidth\columnwidth\hsize\csname@twocolumnfalse\endcsname

\title{Dynamical Mean-Field Solution for a Model of Metal-Insulator Transitions in Moderately Doped Manganites}

\author{Fan Zhong$^{1,2}$,  Jinming Dong$^{1}$,  and Z. D. Wang$^{2}$}

\address{$^{1}$Department of Physics, Nanjing University, Nanjing 210093, People's Republic of China\\
$^{2}$Department of Physics, University of Hong Kong, Hong Kong, People's Republic of China}

\date{\today}

\maketitle

\begin{abstract}
We propose that a specific spatial configuration of lattice sites that energetically favor {\it 3+} or {\it 4+} Mn ions in moderately doped manganites constitutes approximately a spatially random two-energy-level system. Such an effect results in a mechanism of metal-insulator transitions that appears to be different from both the Anderson transition and the Mott-Hubbard transition. Correspondingly, a disordered Kondo lattice model is put forward, whose dynamical mean-field solution agrees reasonably with experiments. 
\end{abstract}


]

% \narrowtext

The discovery of ``colossal'' magnetoresistance (CMR) has stimulated a renaissance of interest in doped rare-earth manganese oxides because of their promising practical applications and their similarity to the cuprate superconductor. \cite{cmr} Although great efforts have been devoted to this system, there is still no consensus on its mechanism so far. In particular, why and how the metal-insulator transitions (MIT's) occur in this system for a moderate doping range are yet to be clearly understood.

Various types of phase transitions are displayed in the perovskite-type manganites of $R_{1-x}A_x$MnO$_3$, with $R$ a trivalent rare-earth element and $A$ a divalent alkaline earth ion. It is well known that the Mn $3d$ levels split in an approximately octahedral ligand into an $e_g$ doublet and a $t_{2g}$ triplet; the former is mobile and the latter half-filled $t_{2g}$ levels are believed to form localized spins of $S=3/2$. The parent compound $R$MnO$_3$ is insulating and antiferromagnetic. Doping of $A^{2+}$ ions creates Mn$^{4+} (t_{2g}^3)$ holes in a Mn$^{3+} (t_{2g}^3 e_{g}^1)$ matrix; and a sufficient number of holes render the mixed-valence compound metallic and paramagnetic or ferromagnetic depending on the temperature. Besides these doping-induced MIT, there is another MIT accompanying the ferro- to paramagnetic phase transition as the temperature is raised in the most interesting doping range of $0.2 \alt x \alt 0.5$. The critical temperature of this transition varies with an applied magnetic field, resulting in the CMR effect. \cite{cmr,tokura94} Lattice effects \cite{hwang,billinge} and even structural phase transitions \cite{magstr} also interplay, making the enigmatic phenomena further intricate.

The mechanism of the CMR effect is yet controversial. It was originally believed to be mediated by the so-called ``double exchange'' since the large ferromagnetic Hund's rule coupling tends to align all the $d$ spins as the mobile $e_g$ electrons hop between the Mn ions. \cite{de} Although the original model of double exchange in the limits of infinite spatial dimensions and classical spins was nicely pursued by Furukawa \cite{furukawa}, Millis and coworkers \cite{millis95} argued that models involving only double exchange yield results such as the magnetic transition temperature $T_c$ and the resistivity above $T_c$ that deviate from the experiments by orders of magnitude. Accordingly they proposed that dynamical Jahn-Teller (JT) effect was crucial. Yet an anomalously large electron-phonon coupling of the JT origin is used to give a qualitative resemblance to the experiments \cite{millis96}. Objections to such vibronic models have also been suggested, emphasizing the effects of interference and localization on the electron scattering \cite{varma}.

Looking into the electrical, magnetic and structural phase transitions, we note the following important facts. The doping-induced MIT can appear in both the para- and ferromagnetic phases \cite{schiff}, and a higher critical doping concentration $x_c$ is needed in the former. It is noted that a paramagnetic metal has only been observed so far in {\em single\/} crystals of La$_{1-x}$Sr$_{x}$MnO$_3$, \cite{tokura94} whose large tolerance factor ($r_{R{\rm -O}}/\sqrt{2}r_{{\rm Mn-O}}$, where $r$ is the averaged distance between the two indicated ions.) approaches 1, a perfect size match. Special attentions should be paid to the sensitivity of the observations to such extrinsic factors as preparation conditions \cite{pickett}. 

Accordingly, an essential ingredient is the double exchange leading to the magnetic ordering. This ordering in turn facilitates hopping of the $e_g$ electrons, resulting in a broadening of the electron bandwidth. Another crucial element is the splitting of the $e_g$ bands. In addition to the broadening of these bands by the spin ordering, which gives rise to a higher $x_c$ for the MIT in the paramagnetic phase, they must be broadened by doping as well, so that doping can trigger MIT's. The analysis here is based on a band-closing MIT, but this is evidenced by spectroscopic observations \cite{park}.

The key point now is the mechanism of the splitting. An important fact that is universal in the doped manganites is the presence of the two valence states, Mn$^{3+}$ and Mn$^{4+}$, which can even become ordered for appropriate doping. A salient effect of the doping is to create a specific spatial arrangement (random but quenched) of $A^{2+}$ such that there is a corresponding preferable spatial distribution of lattice sites that energetically favor {\it 3+} and {\it 4+} Mn ions (we call them {\it 3+} and {\it 4+} sites)\cite{note}. Consequently, that an $e_g$ electron hops from a {\it 3+} sites to a {\it 4+} site to form a Mn$^{3+}$ will cost an extra energy $\Delta$, which was estimated to fall within the order of magnitude of the bandwidth even accounting for reasonable screening\cite{coey}. Fluctuations, induced for example by individual environment, around the two energies should be secondary as the two valence states are prominent. Although such doping-induced disorder is likely argued to trigger a disorder-induced Anderson transition, one usually has to invoke the mass-enhancement from the polaronic effects of size mismatch or of spin disorder in order to realize localization\cite{mott}. We shall show below that the main result of the two randomly distributed levels is to split an $e_g$ band into two subbands, which can become overlapped by doping and/or spin ordering. Such a physical picture seems to dominate the systems under consideration that exhibit conclusively band-closing and behave electronically as a linear superposition of its end-point compounds \cite{park}.

To capture the essential physics, we neglect the fluctuations around $\Delta$ and include a diagonal disorder with only a binary-alloy distribution. As a simplified model, we consider only the classical spin limit \cite{furukawa}. Therefore, the Hamiltonian is given by, 
\begin{equation}
H=\sum_{i,\sigma}\varepsilon_{i}c^{\dag}_{i\sigma}c_{i\sigma}-\!\!\!\!\sum_{<i,j>,\sigma}\!\!\! t_{ij}\!\left[c_{i\sigma}^{\dag} c_{j\sigma}+\mbox {h.c.}\right]-J \sum_{i}\mbox{{\boldmath $\sigma$}}_i \cdot {\bf m}_{i}\label{ham},
\end{equation}
where ${\bf m}_{i}= (m_i^x, m_i^y, m_i^z)$ and $| {\bf m}|=1$, $t_{ij}$ stands for the nearest neighbor hopping integral and $J$ the Hund's rule coupling, $c_{i\sigma}^{\dag} (c_{i\sigma})$ creates (destroys) an electron at site $i$ with spin $\sigma$, and the set of random variables $\varepsilon_{i}$ assumes an independent identical distribution $p(\varepsilon)=x\delta(\varepsilon-\Delta/2)+(1-x)\delta(\varepsilon+\Delta/2)$.

The first two terms of Eq.(\ref{ham}) form simply the Anderson model of disorder, while the last two terms represent a ferromagnetic Kondo lattice model. Our combination of these two models is expected to manifest an interplay of disorder, electricity and ferromagnetism. Note that a similar but far more complex model that includes in addition to the two nonequivalent sites, oxygen orbitals as well as strong Coulomb and exchange interactions has been used to calculate the band structure \cite{mazzaferro}.

A remarkable reward of the two-level approximation is that it renders the model analytically solvable in the limit of infinite spatial dimensions $d$, or, within the dynamical mean-field theory \cite{georges}. Although at $d \rightarrow \infty$, this method is unable to capture the effects of Anderson localization, the results obtained are already nontrivial. In this limit, the disordered system reduces to an ensemble of self-consistently determined Anderson impurity models via proper rescaling of the nearest hopping integral $t_{ij} = t/\sqrt{2d}$. The effective bandwidth $t$, to be set to 1 below, fixes the energy scale. The on-site Green's function $G(i\omega_n)$ is calculated exactly from the effective field ${\cal G}_{0}(i\omega_n)$ as
\begin{equation}
\overline{G}(i\omega_n) = \overline{\langle({\cal G}_{0}(i\omega_n)^{-1}-\varepsilon-J {\bf m}\cdot\mbox{{\boldmath $\sigma$}})^{-1}\rangle_{\bf m}},
\end{equation}
where overbars denote an average over disorder, $\langle \cdots \rangle_{\bf m}$ thermal average over ${\bf m}$, \cite{furukawa} and $\omega_n \equiv (2n+1) \pi / \beta = (2n+1) \pi T$ are the Matsubara frequencies. On the Bethe lattice, the self-consistent condition reads \cite{georges}
\begin{equation}
{\cal G}_0(i\omega_n)^{-1}=i\omega_n +\mu-\overline{G}(i\omega_n)/4, \label{sc}
\end{equation}
ensuring $\overline{G}(i\omega_n) = \sum_{\bf k} G({\bf k}, i\omega_n)= \int D(\epsilon)G(\epsilon, i\omega_n)d\epsilon$, where $D(\epsilon) =2 \sqrt{1-\epsilon^2}/\pi$ is the noninteracting density of states on the Bethe lattice, and
\begin{equation}
G(\epsilon,i\omega_n) = \frac {1} {i\omega_n+\mu-\epsilon-\Sigma (i\omega_n)}, \label{g}
\end{equation}
the exact single-particle Green's function, with the self-energy $\Sigma(i\omega_n) = {\cal G}_{0}(i\omega_n)^{-1}-\overline{G}(i\omega_n)^{-1}$. In Eq.(\ref{sc}), $\mu$ is the chemical potential to be determined by the doping $x$,
\begin{equation}
1-x=\sum_{\sigma}\int_{-\infty}^{+\infty} f(\omega) \overline{A}_{\sigma}(\omega) d\omega,
\end{equation}
where $f(\omega) = 1/(\exp(\beta \omega)+1)$ is the Fermi function, and $\overline {A}_{\sigma}(\omega) =-{\rm Im} \overline{G}_{\sigma}(\omega+i0^{+})/ \pi$ the single-particle density of states (DOS). Moreover, the resistivity is given by \cite{furukawa,georges}
\begin{equation}
\rho(T)=\rho_{0} \left(\sum_{\sigma}\int D(\epsilon) d\epsilon \!\int A_{\sigma} (\epsilon, \omega)^{2}\left[-\frac {\partial f(\omega)}{\partial \omega}\right]d\omega\right)\!^{-1}, \label{sigma}
\end{equation}
where $A_{\sigma}(\epsilon, \omega) =- {\rm Im}G_{\sigma} (\epsilon, \omega+i0^{+})/\pi$ and $\rho_{0}$ gives the unit of the resistivity.

For simplicity, only $J \rightarrow \infty$ is considered. Defining $\mu = -J+\delta \mu$ and $\Omega = \omega+\delta \mu$, \cite{furukawa} and taking the direction of the spontaneous magnetization $M$ to be along the $z$-axis, we finally obtain, (setting $\cos\theta=\xi$)
\begin{mathletters}
\label{main}
\begin{equation}
M=(1-x) \int_{-1}^{1} P_{A}(\xi) \xi d\xi +x \int_{-1}^{1} P_{B}(\xi) \xi d\xi, \label{mainm}
\end{equation}
\begin{equation}
P_{A,B}(\xi)=\frac{1}{Z_{A,B}}\exp\left[\sum_{n=-\infty}^{+\infty} \ln(b_{0 n}-b_{1 n} \xi \pm \Delta) \right], \label{mainp}
\end{equation}
\begin{equation}
b_0=2\Omega-\frac{1\!-x}{2}\!\!\!\int \!\frac {P_A( \xi )}{b_0-b_{1} \xi  + \Delta} d\xi -\frac{x}{2}\!\! \int \!\frac {P_B( \xi )}{b_0-b_{1} \xi-\Delta}d\xi, \label{main0}
\end{equation}
\begin{equation}
b_1=\frac{1-x}{2}\int \frac {P_A( \xi ) \xi }{b_0-b_{1} \xi + \Delta}d\xi + \frac{x}{2} \int \frac {P_B( \xi ) \xi } {b_0 - b_{1} \xi - \Delta}d\xi, \label{main1}
\end{equation}
\end{mathletters}
with $\overline{G}_\sigma=4\Omega-2b_0+2\sigma b_1$, ($\sigma=\pm 1$), $Z_{A,B}=\int \exp\left[\sum_n \ln(b_{0 n}-b_{1 n} \xi \pm \Delta)\right] d\xi$, $b_{0 n}\equiv b_{0}(i\omega_n)$ and $b_{1 n}\equiv b_{1}(i\omega_n)$.

For a given splitting $\Delta$, doping $x$ and temperature $T$, the set of Eqs.(\ref{main}), can be solved generally by numerical methods. Several situations can, however, be further treated analytically. These are $T=0$, $\Delta=0$, and $T>T_c$, the magnetic transition temperature just below which $b_1 \neq 0$.

\begin{figure}
% \vspace{2.3in}
%\special{psfile=dos335w.eps angle=90}
\epsfysize 2.3in \epsfbox{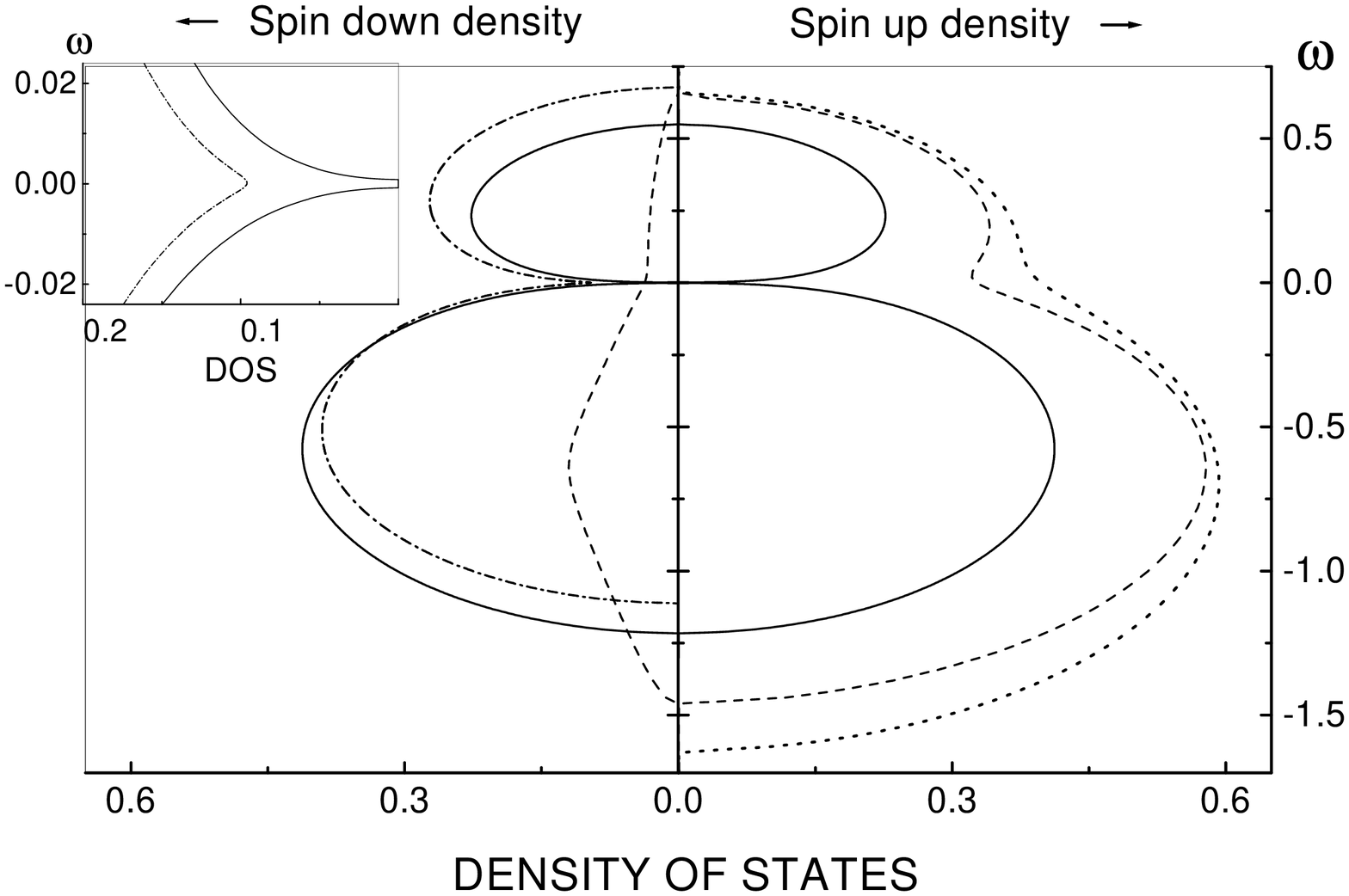}
\caption{Density of states for $x=0.2$ at different temperatures. Solid: $T=0.035$ ($>T_c$); dashed: $T=0.01$ ($<T_c$); and dotted: $T=0$. For comparison, the spin down density for $x=0.3$ at $T=0.035$ ($>T_c$) (dash-dotted) is given in both the main figure and the inset (magnified plot near $\omega=0$)}.
\label{dos}
\end{figure}

For $T>T_c$, no magnetic ordering appears and so $b_1=0$, $P_{A}=P_{B}$ and $M=0$, and $\overline{G}_{\uparrow}=\overline{G}_{\downarrow}\equiv \overline{g} (\omega)$. One immediately obtains from Eqs.(\ref{main}),
\begin{equation}
\overline{g}^{3}-8\Omega\overline{g}^{2}+(16\Omega^2-4\Delta^{2}+2) \overline{g} - 8\Omega+4(2x-1)\Delta = 0,\label{parag}
\end{equation}
which is identical to Eq. (44) of Ref. \onlinecite{vlaming} except that a different energy scale is chosen. For $T=0$, the main contribution to $P_A$ and $P_B$ comes from $\xi=1$. Consequently, $b_1 = 2\Omega-b_0$, so $\overline{G}_{\uparrow}=8\Omega-4b_0$, but $\overline{G}_ {\downarrow}=0$, {\it i.e.}, the electrons are completely polarized. One finds thus a similar equation to Eq.(\ref{parag}) except with different coefficients that lead to a substantial broader bandwidth (Fig.~\ref{dos}). 
For $\Delta=0$, we recover the results of Furukawa \cite{furukawa}. One obtains, for $T>T_c$, $g(\omega)=2\Omega-2i\sqrt{1/2-\Omega^2}$, while at $T=0$, $G_ {\uparrow}(\omega)=2\Omega-2i\sqrt{1-\Omega^2}$. Thus for moderate doping, there are only partly filled bands so that no insulating behavior exists at all.

With these known cases as references, it is not difficult, albeit tedious, to solve Eqs.(\ref{main}) self-consistently. Here we shall only present the results for the DOS and $\rho(T)$. We choose reasonably the effective bandwidth to be 1.12eV and $\rho_0 =10^{-3}$ ($\Omega$cm)$^{-1}$. \cite{furukawa} Thus a splitting $\Delta=0.67$ corresponds to 0.75eV and $T=0.01$ to 130.0K. 

Figure~\ref{dos} displays the DOS for several values of the parameters, manifesting various possible transitions. For each orientation of the localized spins, the $e_g$ band splits, due to the Hund coupling, into a high-energy antiparallel band and a low-energy parallel band; only the latter is relevant here, as the separation is of the order of $J$. Such a splitting is similar to the Hubbard coulomb correlation. The new point here, however, is that for a nonzero $\Delta$, each band splits further into two bands. From the spin ``down'' densities at $T=0.035(>T_c)$ (identical to the ``up'' ones) of $x=0.2$ and $x=0.3$ for $\Delta=0.67$, it is seen that the spectral weight transfers from the lower filled electron band to the upper empty hole band as $x$ increases. As a result, when the critical doping $x_c$ is reached, the two bands overlap, leading to a doping-induced MIT.

At $T=0.01$, lower than $T_c$, the magnetic ordering renders the spectral weight for the two spin orientations unequivalent (dashed lines). More importantly, it substantially broadens the bandwidth, so that the two splitted bands of $x=0.2$ at $T>T_c$ now overlap. Consequently, an insulator-metal transition occurs accompanying with the para- to ferromagnetic transition. Moreover, as the bands broaden in the ferromagnetic phase, $x_c$ gets smaller for the same $\Delta$. 

On the other hand, as $\Delta$ increases, a larger $x_c$ is required. For a sufficiently large $\Delta$, no doping-induced MIT will be present in the paramagnetic phase. This is likely the case for systems with a more distorted lattice or a smaller tolerance factor. When $\Delta$ becomes so large that even for $T \ll T_c$, the two bands are still splitted, a transition then occurs from a paramagnetic insulator to a ferromagnetic insulator, or a canted antiferromagnetic insulator due to the antiferromagnetic superexchange coupling between the localized spins.

\begin{figure}
%\vspace{2.3in}
\epsfysize 2.3in
\epsfbox{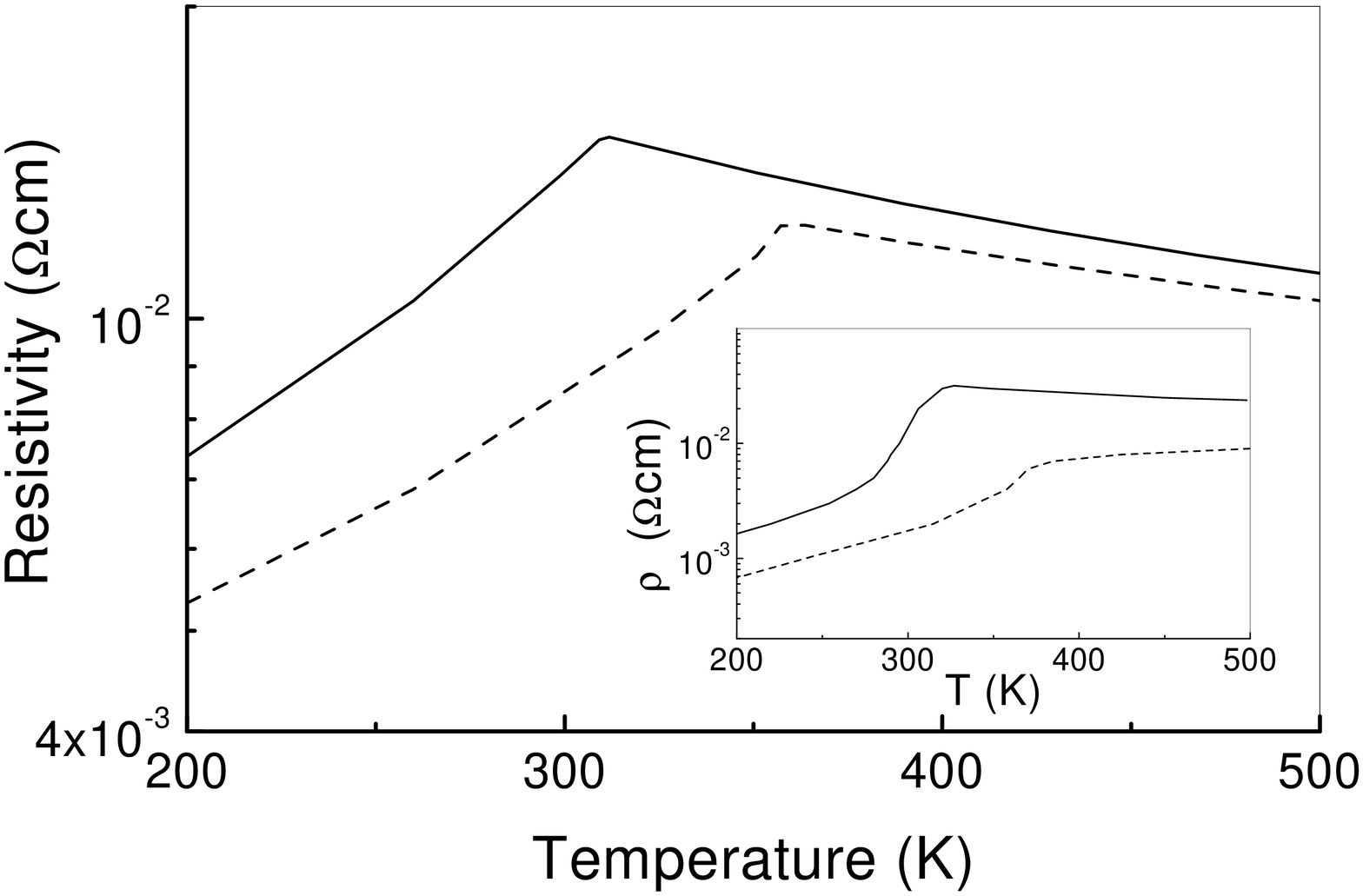}
\caption{Resistivity vs temperatures. Experimental data from [2] are shown in the inset. Solid lines: $x=0.2$; and dashed: $x=0.3$.}
\label{rho}
\end{figure}

The resistivity is shown in Fig.~\ref{rho} with experimental data taken from Ref.~\onlinecite{tokura94}. The agreements, including the positions of the peaks signaling $T_c$, are remarkable by noticing that only a {\em single\/} set of $\Delta$ and $\rho_0$ has been chosen and several limits have been made in such an extremely simplified theory. Note that although for $x=0.3$ and $T > T_c$ the bands have already overlapped, the resistivity still exhibits somehow an insulating behavior $(d\rho/dT<0)$. This ``pseudogap'' behavior has also been found in Ref.~\onlinecite{millis96}, namely, as soon as a dip, rather than a real gap, develops at $\omega=0$ in the spectral function, the resistivity starts to rise with decreasing temperature.
Actually, $\Delta$ and $\rho_0$ depend in general on $x$. So we may readily choose other sets of the parameters to produce a better agreement with experiments, but this is trivial, as Fig.~\ref{rho} has already touched the essence. The reason of the deviations from experiments is conceivable. As has been pointed out above, once the spins are in order and hence the bands broaden below $T_c$, the bonding strengthens, leading to a more compact or less distorted lattice. Consequently, the effective bandwidth $t$ increases below the transitions, giving rise to the more drastic drops of the resistivities observed across $T_c$. Therefore, a more realistic description of the CMR should include the coupling to the lattice degrees of freedom in order to ``dress'' the ``bare'' $t$ and also $\Delta$, so that they can vary with the transitions. 

In summary, we have put forward a theory that explains naturally the doping-induced metal-insulator transitions in both the para- and ferromagnetic phases, as well as the magnetic phase transitions, with the results of doping- and temperature-dependent resistivities being in reasonable agreement with experiments. Essential in the theory is a random lattice of two-energy levels that represent the relative energies of the two species of sites that energetically favor {\it 3+} and {\it 4+} Mn ions created by doping. So our theory is more appropriate for moderately doping \cite{note1}.  The central physical picture is that the Hund's-rule splitted $e_g$ band is further splitted on the two levels into a filled electron and an empty hole bands, which can become overlapped by doping and/or spin ordering. Such a mechanism of electronic MIT appears to be different from the two usual classes \cite{belitz}, the Mott-Hubbard and the Anderson transitions, since the former is due to coulomb correlation, and the latter is realized by sweeping the Fermi energy across the mobility edge. Further, it may be expected to shed light to some other multicomponent systems.
Although our model is extremely simplified, the physical picture for the metal-insulator transitions is simple but clear, and the theoretical results are in reasonable agreement with experiments. The usual Anderson transition should of course be investigated. Further extensions of the theory are almost transparent. One may study, for example, effects of finite $J$, vibronic coupling, multibands, and the Coulomb correlation, to list but a few. Applications to other problems are also attracting.

We thank Prof.~Zhengzhong~Li for his illuminating discussions. One of us (FZ) acknowledges the support of the Postdoctoral Science Foundation of China. This work was supported by the NSF of China, No.19677202, and a CRCG grant at HKU.


\begin{references}
\bibitem{note1}For undoped LaMnO$_3$, additional anisotropic interactions should be included for a more complete description.
\end{references}
\end{document}